\documentclass[conference]{IEEEtran}
\IEEEoverridecommandlockouts
\usepackage{cite}
\usepackage{amsmath,amssymb,amsfonts}
\usepackage{algorithmic}
\usepackage{graphicx}
\usepackage{textcomp}
\usepackage{xcolor}
\usepackage{booktabs}  
\usepackage{amsmath}   

\def\BibTeX{{\rm B\kern-.05em{\sc i\kern-.025em b}\kern-.08em
    T\kern-.1667em\lower.7ex\hbox{E}\kern-.125emX}}
\begin{document}

\title{Shrinking: Reconstruction of Parameterized Surfaces from Signed Distance Fields\\
}

\author{\IEEEauthorblockN{1\textsuperscript{st} Haotian Yin}
\IEEEauthorblockA{\textit{Dept. of Computer Science} \\
\textit{New Jersey Institute of Technology}\\
Newark NJ, USA \\
hy9@njit.edu}
\and
\IEEEauthorblockN{2\textsuperscript{nd} Przemyslaw Musialski}
\IEEEauthorblockA{\textit{Dept. of Computer Science} \\
\textit{New Jersey Institute of Technology}\\
Newark NJ, USA \\
przem@njit.edu}
}

\maketitle

\begin{abstract}
We propose a novel method for reconstructing explicit parameterized surfaces from Signed Distance Fields (SDFs), a widely used implicit neural representation (INR) for 3D surfaces. While traditional reconstruction methods like Marching Cubes extract discrete meshes that lose the continuous and differentiable properties of INRs, our approach iteratively contracts a parameterized initial sphere to conform to the target SDF shape, preserving differentiability and surface parameterization throughout. This enables downstream applications such as texture mapping, geometry processing, animation, and finite element analysis. Evaluated on the typical geometric shapes and parts of the ABC dataset, our method achieves competitive reconstruction quality, maintaining smoothness and differentiability crucial for advanced computer graphics and geometric deep learning applications.
\end{abstract}


\section{Introduction}
Implicit Neural Representations (INRs)~\cite{deepsdf} have become popular 3D models in computer graphics, with applications in scientific simulation, photogrammetry, generative modeling, and inverse physics~\cite{implicitsurvey}. INRs encode continuous signals via neural networks that map spatial coordinates to signal values~\cite{meshsdf}, offering advantages like efficient storage, smooth interpolations, and differentiable features, surpassing traditional grid-based methods~\cite{geoprocessing}. Among them, Signed Distance Fields (SDFs) are widely used to implicitly define surfaces~\cite{deepsdf}. However, SDFs pose challenges for tasks needing explicit surface representation, like direct rasterization. Traditional reconstruction methods such as Marching Cubes (MC)~\cite{mc} and its variants~\cite{NDC,flexicube} divide space into cubes to identify surface intersections. While effective, these approaches generate discrete meshes that fail to leverage the continuity and differentiability of INRs.

\begin{figure}[t]
    \centering
    \includegraphics[width=0.49\textwidth]{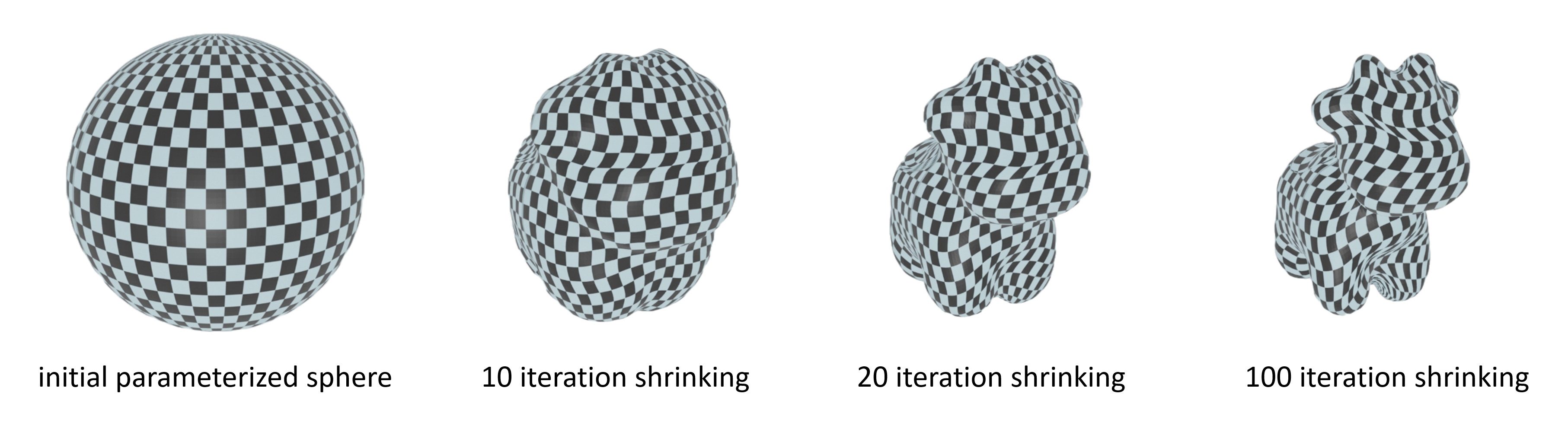}
    \caption{We introduce a shrinking method, iteratively morphing a parameterized sphere to extract an implicitly represented surface, maintaining surface parameterization. From left to right, it show the shrinking process through iteration through the Spot geometry model.}
    \label{para_spot} 
\end{figure}

One way to preserve these properties is surface parameterization, which involves a bijective mapping between a 2D domain and a 3D surface~\cite{spherepara_gu}. Parameterization maintains surface continuity, enabling complex deformations required in applications like virtual reality and medical imaging. Most existing parameterization methods build from discrete meshes~\cite{seamlesspara}, parameterizing discrete vertices and using barycentric interpolation, resulting in surfaces that are not differentiable.

We propose a novel shrinking method to directly extract explicit, differentiable, parameterized surfaces from Neural SDFs, bypassing the need for intermediate discrete meshes and preserving differentiability from the INR to the parameterized surface. Inspired by~\cite{shrinkwrap}, our method iteratively contracts a parameterized sphere to match the target shape, guided by signed distance values through gradient descent. Each step involves remeshing to maintain uniform distribution, ensuring surface continuity and smooth parameterization.

Our experiments demonstrate that this approach not only generates differentiable parameterizations but also achieves competitive reconstruction quality compared to mainstream methods. Our contributions include: (1) Introducing a shrinking-based method for extracting high-quality meshes from Neural Implicit Representations. (2) Achieving global surface parameterization, expanding INR applicability. (3) Enabling spherical parameterization for complex models.

\section{Related Work}
\subsection{Implicit Neural Representations (INRs)}

An implicit neural representation (INR) is a way of encoding a signal $s(x): \mathbb{R}^m  \rightarrow  \mathbb{R}^n$ by means of fitting a neural network $f_\theta(x)$ (of the same domain and codomain) parameterized by $\theta \in \mathbb{R}^p$ on a dataset $\mathcal{X} = \{x_i\}$ of sample coordinate points such that we minimize signal reconstruction loss~\cite{deepsdf}:
$$\mathcal{L}(\theta) = \mathbb{E}_{x \sim \mathcal{X}}[(f_\theta(x) - s(x))^2] + \sum_i \lambda_i R_i(f_\theta, \cdot)$$
Assuming that the sample points $x$ are distributed uniformly within the sample set $\mathcal{X}$, it represents a regression. In finding an optimal model $f_\theta$, we have a memory-efficient, and flexible continuous encoding of the signal $s(x)$. 

\subsection{Signed Distance Fields}

A Signed Distance Field (SDF) is a scalar field that defines a surface by assigning each spatial coordinate a distance value to the nearest point on the surface, with the sign indicating whether the point is inside (negative) or outside (positive) the watertight surface. 
The surface is implicitly defined by the isosurface condition \( SDF(\cdot) = 0 \), representing the object's shape and spatial positioning~\cite{trasdfssd}.

Neural networks have become prominent for generating implicit representations like SDFs due to their ability to effectively address inverse problems in model reconstruction. These networks encode the shape into their parameters, optimized over sampled input points from a shape or latent space. Techniques using Multi-Layer Perceptrons (MLPs) generate a reconstruction domain that is compared to the actual sensor domain, with a loss function derived for iterative refinement through gradient descent~\cite{nerf}.

Park et al.~\cite{deepsdf} introduced an auto-decoder optimizing the latent code of a shape using Maximum a Posteriori (MAP) estimation, representing shapes as continuous SDFs. This enables SDF estimation at any point, aiding in applications like shape interpolation and optimization.

Following Park's work, advancements have focused on enhancing loss functions with gradients and second-order derivatives~\cite{netgropp2020}, and modifying network architecture with detail-enhancing elements or sophisticated hidden layers~\cite{netwilliams21kernal}. 
These advancements significantly improve neural network methodologies for surface approximation and representation.

\subsection{Surface Reconstruction}
\subsubsection{Spatial Decomposition}
The MC algorithm revolutionized 3D surface reconstruction by dividing a scalar field into a grid of cubes, approximating the surface within each cube with triangles formed by linearly interpolating edge values based on scalar data. 
$$u_e=\frac{x_a \cdot s\left(x_b\right)-x_b \cdot s\left(x_a\right)}{s\left(x_b\right)-s\left(x_a\right)}$$
Liao~\cite{deepMC} and Remelli~\cite{meshsdf} highlighted a singularity when $s\left(v_a\right) = s\left(v_b\right)$, potentially disrupting differential optimization, though this issue is often avoided in practice~\cite{deepmtet}. 
The MC method treats each cube independently, resulting in the reconstructed geometry losing contextual information beyond the boundaries of individual cubes. This limitation lost the integrity of the surface, and hinders the mesh's ability to conform to sharp features and inevitably results in the formation of low-quality sliver triangles, particularly when the isosurface closely approaches a grid vertex. Recent methods like Recent methods like Neural Marching Cubes (NMC)~\cite{NMC}, Neural Dual Contouring (NDC)~\cite{NDC}, and FlexiCubes~\cite{flexicube} enhance mesh quality by integrating neural networks for local adjustments, though they still lack global surface parameterization.

\subsubsection{Surface Tracking}
Surface tracking methods leverage neighboring surface samples for isosurface extraction. A notable approach, Marching Triangles by Hilton et al.~\cite{marchingtri}, uses iterative triangulation under a Delaunay constraint, generating new vertices by expanding triangle edges from an initial point. Later advancements introduced adaptivity\cite{surtrack01_2} 
and sharp feature alignment~\cite{surtrack02}. However, implementing gradient-based mesh optimization in this framework is challenging due to the need to differentiate through the discrete, iterative update process.

\subsubsection{Shrink Wrapping}
Traditional Shrink Wrapping or inflation methods utilizes strategies like contracting a spherical mesh~\cite{shrinkwrapvan2004} or inflating critical points~\cite{shrinkwrapbottino1996}
to align with the isosurface. These methods are inherently limited to specific topological scenarios and often necessitate manual selection of critical points  to accommodate arbitrary topologies. Furthermore, the differentiation through the shrink wrapping process is complex, making these methods less conducive to gradient-based optimization techniques.

\subsection{Surface Parameterization}
Surface parameterization research has primarily focused on achieving global seamless parameterizations for mesh structures. Disk-like surfaces are ideal candidates for continuous, injective parameterization, either locally or globally. For surfaces with arbitrary topology, a common strategy is to segment them into disk-like patches, applying parametric transitions across cuts to ensure seamlessness~\cite{paraquad19}~\cite{paraseamless12}. 
Recent advances in seamless parameterization~\cite{seamlesspara} have focused on cone structures, providing a simpler constructive proof for the existence of seamless surface parameterizations and translating it into practical algorithms. For any set of topologically valid parametric cones with specified curvature, a global seamless parameterization can be achieved, with one well-known exception. Shen et al.~\cite{holonomypara} notably enhance this by modifying the loop system to achieve a topologically equivalent state, allowing comprehensive control over map topology in the parameterization process.

\begin{figure}[h]
  \centering
  \includegraphics[width=4.5cm]{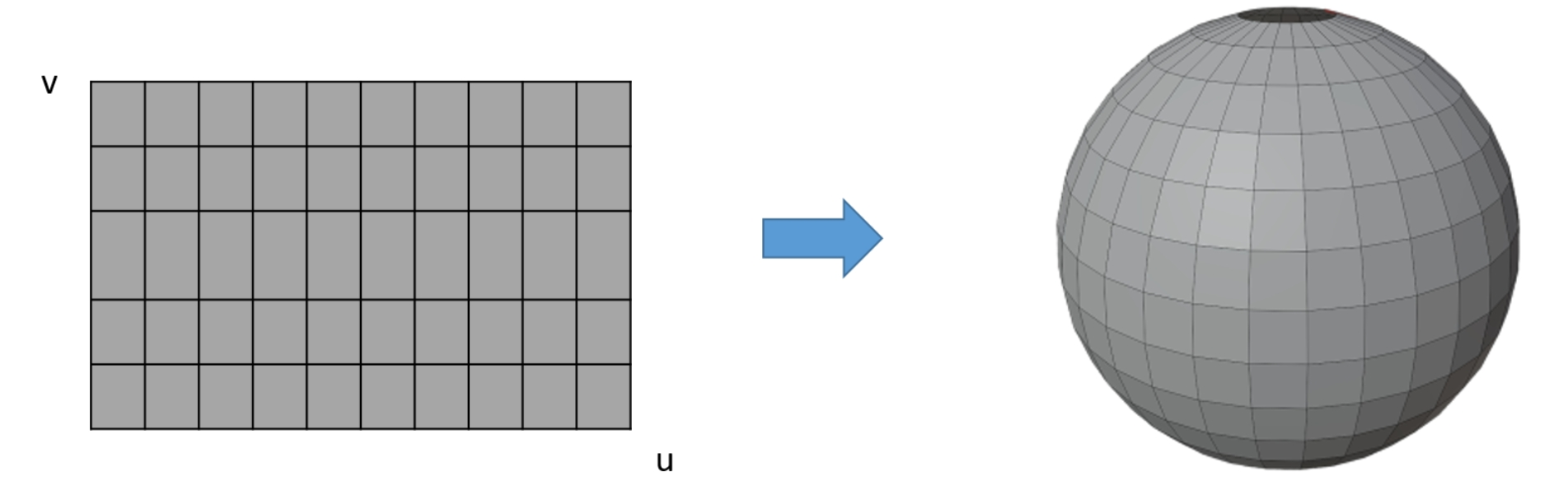}
  \caption{Sketch map for anchor points base on spherical coordinates, we then get a initial parameterized sphere}
  \label{anchor}
\end{figure}
\label{ballsketch}

\section{Methodology}
In this section, we introduce a method for reconstructing parameterized curves from implicit representations. The core technique iteratively contracts an initialized circle to align with the target shape, represented as the iso-surface. Our key contribution is a scaled gradient optimization for curve updates, minimizing the absolute value of signed distance to ensure convergence. To handle concave regions and reduce parameterization distortion, we integrate remeshing.

\subsection{Shape representation}
We represent a generic closed shape \( S \) using a signed distance function (SDF), defined as \( s: \mathbb{R}^2 \rightarrow \mathbb{R} \) in 2D and \( s: \mathbb{R}^3 \rightarrow \mathbb{R} \) in 3D. The SDF maps each point \( x \) in space to a scalar value that indicates the shortest Euclidean distance to the surface of \( S \). This value is positive if \( x \) is outside \( S \), and negative if inside. Thus, the shape \( S \) is defined as the zero-level-set of the SDF, where the function equals zero. The formal SDF definition is:

\small
$$
s(x) = 
\begin{cases} 
  \rho(x, S), & \text{if } x \text{ is outside } S \\
  -\rho(x, S), & \text{if } x \text{ is inside } S
\end{cases}
$$
\normalsize
Here, $\rho(x, S) =\min _{\mathbf{y} \in S} d(\mathbf{x}, \mathbf{y})$  represents the Euclidean distance between the point \( x \) and the nearest point on the surface of the shape $S$.

In order to accurately compute the Signed Distance Function (SDF) for complex shapes, we employ the methodology delineated in DeepSDF ~\cite{deepsdf}, which leverages a neural network architecture in conjunction with a latent space representation $\mathbf{z}$ to model the SDF of a shape. This approach utilizes an auto-decoder framework, wherein the network is trained categorically on a variety of shapes. This training results in a generalized neural network model capable of representing a wide array of shapes, in addition to a unique latent vector. The Multi-Layer Perceptron network $f_\theta$ is trained to approximate $s(x)$ over the shape $S$ by minimizing~\cite{meshsdf}

\small
\begin{equation}\nonumber
\begin{aligned}
\mathcal{L}_{\text{sdf}}\left(\left\{\mathbf{z}_S\right\}_{S \in \mathcal{S}}, \theta\right) = \sum_{S \in \mathcal{S}} \frac{1}{\left|X_S\right|} \sum_{\mathbf{x} \in X_S} \left| f_\theta\left(\mathbf{x}, \mathbf{z}_S\right) - s(\mathbf{x}) \right| \\
+ \lambda_{\text{reg}} \sum_{S \in \mathcal{S}} \left\|\mathbf{z}_S\right\|_2^2
\end{aligned}
\end{equation}
\normalsize
where $\mathbf{z}_S \in \mathbb{R}^Z$ is the latent vector encoding the geometric characteristics of the shape $S$, $\theta$ denotes the network parameters, The set $X_S$ represents a collection of points that are sampled from the original shape $S$ to train the network,  and the regularization term, scaled by $\lambda_{\text {reg }}$ is incorporated to prevent over-fitting.

\subsection{Shape Shrinking}\label{shrinking}

\subsubsection{Shrink Wrapping}
We conceptualize the process as iteratively contracting a watertight initial surface $C$ until it precisely conforms to the contours of the implicit target shape $S$. The primary objective of this approach is to minimize the unsigned area enclosed between the surface $C$ and the shape $S$, ideally reducing it to a value approaching zero. This minimization strategy can be mathematically articulated as the summation of the shortest Euclidean distances from each point on the surface $C$ to the nearest point on the shape $S$. Formally, this is equivalent to computing the integral of the absolute value of the signed distance function over the surface $C$: 

\small
\begin{equation}
\begin{aligned}\nonumber
J(C) & =\frac{1}{|C|} \oint_C \lvert s(x) \rvert dx \\
& =\frac{1}{|C|} \oint_C\lvert s(x(u,v))\rvert  x(u,v) \, dudv
\end{aligned}
\end{equation}
\normalsize
where $x(u,v)$ is the parameterized points on the surface $C$. Make it discrete, we have: 

\small
$$J(C)=\frac{1}{|C|} \lim _{n,m \rightarrow \infty}\sum_{i=1}^n \sum_{j=1}^m\rvert s\left(x(i,j)\right)\rvert (\Delta x)^2$$
\normalsize

By sampling points $x_C$ on the curve $C$, we have the discrete approximation of the loss function

\subsubsection{Iterative Shrinking by Scaled Gradient}
We initiate with a sphere using spherical coordinates and uniformly sample points on a grid in UV space. We iteratively contract the sphere until it conforms the shape \( S \). 
Based on MeshSDF~\cite{meshsdf}, a point $\mathbf{v}$ lying on its iso-surface $S = \{ \mathbf{q} \in \mathbb{R}^3 | s(\mathbf{q}) = 0 \}$, the evolution of how the iso-surface move when $s$ undergoes a perturbation $\Delta s$ can be proved as 

\small
$$
\frac{\partial \mathbf{v}}{\partial s}(\mathbf{v}) = -\mathbf{n}(\mathbf{v}) = -\nabla s(\mathbf{v}) \,.
$$
\normalsize
In this expression, $ -\nabla s(\mathbf{v}) $ represents the negative gradient of the SDF at $ \mathbf{v} $. This formulation implies that the shrinking of the iso-surface, in response to the SDF perturbation, progresses in a direction opposite to the SDF's gradient.

To  minimize the function $J$, we employ a scaled gradient optimization technique. The gradient, normalized and then scaled by the signed distance value, is used to iteratively update the curve's position, with each iteration involving a step scaled by the product of the scaled gradient and a predefined step length. We can shrink the surface by updating the sampled points $x \in X_C$ 
\small
$$x_{k+1}=x_k+\frac{\nabla s(x_k)}{\| \nabla s(x_k)\| + \epsilon} \cdot s(x_k) \cdot t$$
\normalsize
where \(k\) denotes the \(k\)-th step, \(t\) denotes the step size, and \(\epsilon\) is a small constant added to prevent division by zero.

\subsection{Shrinking Parameterization in 3D}
In our implementation, we first train the neural network to model the target surface. We then use a spherical shape that gradually morphs to match the target watertight surface. Uniform spherical coordinate-based sampling establishes anchor points for initial parameterization, and an innovative resampling technique refines the point distribution during the iterative shrinking process, ensuring uniform coverage and avoiding local minima.


\subsubsection{Network Training}
We employ a network to implicitly represent the 3D surface using the DeepSDF~\cite{deepsdf} method. Training samples are generated based on the signed distance within a point set:
\small
$$ \text{TRAIN} := \{((x, y, z), s) : \text{SDF}(x, y, z) = s\} $$
\normalsize
Point selection is conducted through uniform sampling around the figure, focusing on the signed distance function.

\subsubsection{Build initial sphere by spherical coordinate} 
We initialize the shrinking process with a sphere and use spherical coordinates to uniformly sample points on its surface. A point on the sphere is defined by \((r, \theta, \phi)\), where \(r\) is the radius, \(\theta\) is the azimuthal angle \((0 \leq \theta < 2\pi)\), and \(\phi\) is the polar angle \((0 \leq \phi \leq \pi)\). To achieve uniform distribution, \(\theta\) and \(\phi\) are discretized into \(N_{\theta}\) and \(N_{\phi}\) segments, respectively, allowing control over the mesh resolution. Each point's coordinates are given by: 
\small
\[
x = r \sin(\phi) \cos(\theta), \quad y = r \sin(\phi) \sin(\theta), \quad z = r \cos(\phi),
\]
\normalsize
where \(\theta_i = \frac{2\pi i}{N_{\theta}}\) and \(\phi_j = \frac{\pi j}{N_{\phi}}\). This method ensures evenly spaced points on the sphere, used as anchor points for shrinking (Figure ~\ref{anchor}).



\subsubsection{Shrinking iteratively through the sphere}
\begin{figure}[t]
    \centering
    \includegraphics[width=0.49\textwidth]{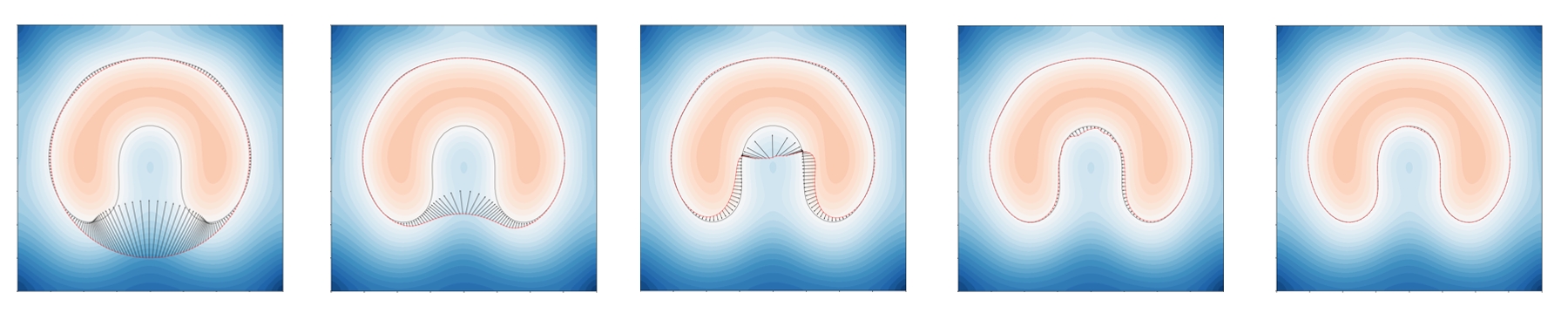}
    \caption{From left to right is the sketch map of shrinking in a 2D scenario. The signed distance field implicitly encodes the target shape in the 2D plane, with blue indicating positive signed distances and red indicating negative. The process begins with an initial circle, which morphs into the target shape. Parameterized anchor points are set on the circle and gradually shrink by following the scaled gradient of the signed distance field until they conform to the target shape. Momentum processing is also incorporated in the 2D scenario when shrinking.}
    \label{shrink}
\end{figure}

Following the method outlined in Section~\ref{shrinking}, we iteratively shrink our anchor points. For each anchor point \(x\) in the set \(X_C\), we update the points by following the gradient direction of our shrinking process. We scale our step size based on the signed distance value, which allows for larger initial steps that progressively focus on finer shape details as the shrinking continues (Fig~\ref{shrink}). We can demonstrated that our shrinking process converges to the target shape.


\subsection{Remeshing}
\subsubsection{Resampling} 
Our method employs resampling techniques to adapt to the parameterization of complex shapes. In the original method without resampling, two issues may arise when dealing with intricate geometries: the final convergence in the shrinking process might settle in local minima, making it challenging to handle concave shapes; and the uneven distribution of anchor points during shrinking can lead to reduced precision in certain areas of the shape. To address these issues, we have designed a resampling trick for each iteration of the shrinking process.

We utilize curve resampling to achieve surface resampling. For both horizontal and vertical directions on the parameterized plane, we map anchor points along each line in these two directions to the shrinking surface, forming a polyline. The anchor points on this polyline are then adjusted to ensure uniform distribution along it. In each step of the shrinking process, we alternate resampling in both directions, ultimately attaining the target shape. This approach avoids complex calculations on 2D surfaces and has shown effective results. Additionally, it preserves the positional order of points in the parameterization, thus preventing overlap in parameterization.

\subsubsection{Parameterization}
By establishing a one-to-one correspondence between points before and after the shrinking process, we naturally achieve spherical parameterization. The points on the sphere are mapped to a plane using spherical coordinates, resulting in the shape's parameterization on the plane. Fig~\ref{para_spot} shows the mapping of a checkerboard pattern across different iterations. In our parameterization, the UV coordinates of anchor points on the plane are determined by their spherical coordinates, specifically defined by the initial mesh coordinates \(\theta_i\) and \(\phi_j\) on the sphere. As the shrinking process progresses, the 3D coordinates of these anchor points morph into the target shape, producing a parameterization of the final surface.

\subsubsection{Overlap Prevention in Parameterization Mapping}
In our approach, the 2D parameterization map and vertex connectivity are pre-generated using spherical coordinates. Our task primarily involves altering the positions of anchor point vertices on the sphere to gradually wrap the original surface during iteration. To prevent overlap during parameterization, it is essential to ensure that the vertices' coordinates of anchor points do not overlap during contraction. Our bidirectional resampling process effectively guarantees this. Through resampling, we ensure that vertices in both directions remain sequentially arranged during shrinking, maintaining consistency in the mapping between two and three dimensions. Thus, our model inherently avoids overlap by design.

\section{Experiments}
\subsection{Dataset}
Our experiment data is divided into two parts. First, for qualitative results, we use typical shapes like Spot and Duck, rendering them with a checkerboard pattern to demonstrate our parameterization results. Second, for quantitative results, we train our shrinking method using a selection of data from the ABC dataset~\cite{abcdata}, following the protocols in NDC~\cite{NDC}. The ABC dataset comprises watertight triangle meshes of CAD shapes, known for their rich geometric features. For our experiments, we use only the first chunk of the ABC dataset and select shapes with genus 0.

During data preprocessing, we sample data over a $32^3$ grid to obtain SDF values based on the original mesh. These sampled data points serve as our training data for coordinate-based network. The original mesh is also used as the ground truth for evaluating our reconstructed meshes. To extract SDFs from the original shape meshes, we follow the methodology outlined in~\cite{mesh2sdf22}.

\subsection{Training}
We employ the method outlined in DeepSDF \cite{deepsdf} for training, which utilizes a latent vector and a network tailored to each category to reconstruct the shape. Our network is configured with SIREN activation functions \cite{Sitzmann2000_siren}.


Our approach utilizes a four-layer fully connected neural network as the decoder, featuring a hidden width of 128 and a learning rate of 1e-3. The network is trained over 4,000 epochs, achieving an error rate that approaches zero. During the shrinking phase, we initialized our parameterized sphere with a resolution of 100x200 to facilitate our checkerboard rendering process. We maintained a step size of 0.2 during shrinking, ensuring controlled and precise contraction of the sphere towards the target shape.

\subsection{Evaluation}\label{sec:evaluation} We adopted a subset of the evaluation metrics from NDC~\cite{NDC} to assess our reconstruction method. The efficacy and quality of surface reconstruction were quantitatively evaluated by uniformly sampling 10,000 points on the surface of both the ground truth shape and the predicted shape. We computed a suite of metrics to assess the reconstruction quality, including Chamfer Distance (CD), which evaluates the overall quality of a reconstructed mesh, and Normal Consistency (NC), which measures the accuracy of the surface normals. These metrics are effective in identifying significant errors, such as missing parts of the mesh. We benchmarked our method against the MC algorithm, ensuring that both methods had comparable computational complexity for a fair comparison. Specifically, our reconstructions used a resolution of 100x200, while MC was evaluated at a resolution of \(27^3\).

\section{Results}

\subsection{Qualitative results}
\begin{figure}[h]
  \centering
  \includegraphics[width=0.98\columnwidth]{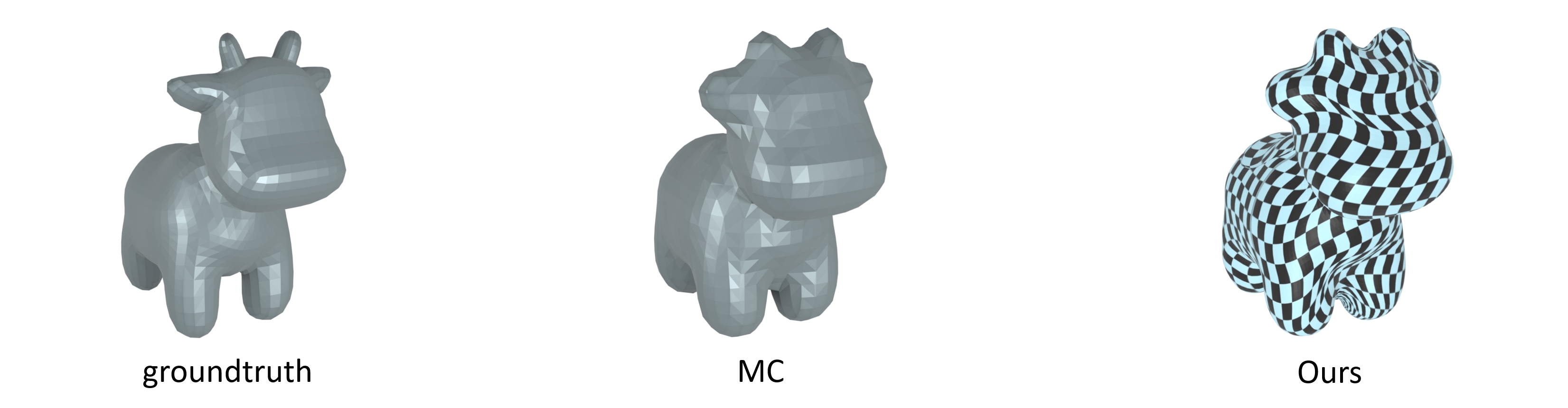}
  \includegraphics[width=0.98\columnwidth]{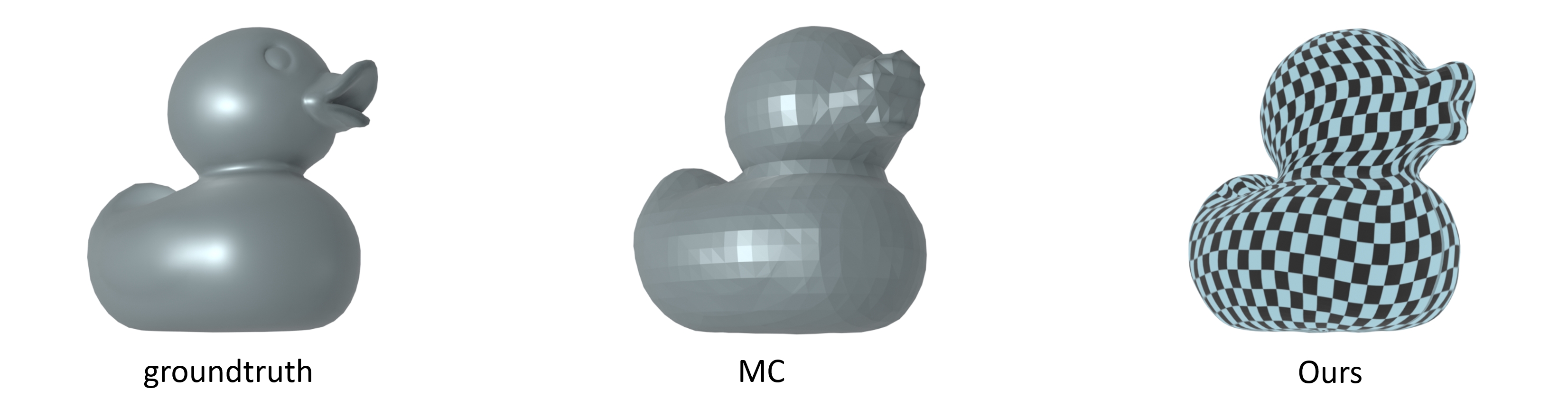}
  \includegraphics[width=0.98\columnwidth]{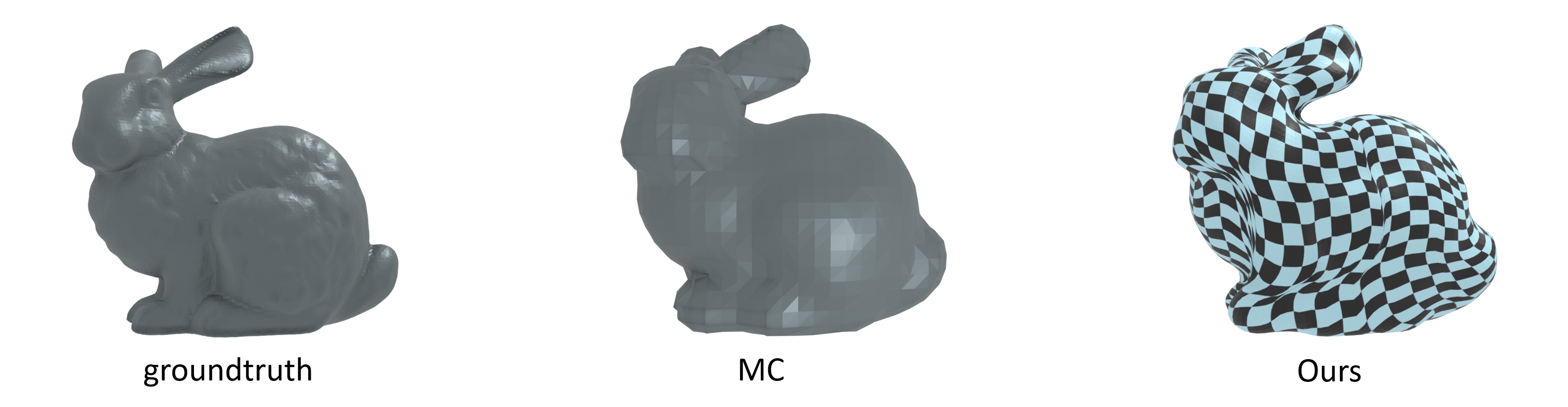}
  \caption{Qualitative comparisons of 3D surface reconstruction on Spot, Bunny and Duck model with MC and our shrinking parameterization method. Please note, we intentionally use a low-resolution volume discretization for the MC result in order to match the size of the resulting triangles to that of our results (cf. Section~\ref{sec:evaluation}). }
  \label{para_checkerboard}
\end{figure}

We render the example shapes, reconstructed using both the MC method and our method, and compare them with the ground truth. As shown in Fig ~\ref{para_checkerboard}, our method performs comparably to the MC method with the same computational complexity while maintaining excellent surface differentiability. This demonstrates our method’s capability reconstructing smooth, high-quality surfaces encoded by neural networks. We also provide an evaluation of our reconstruction capability.

More importantly, we are the first to achieve parameterization directly from implicit neural signed distance fields, whereas other methods like MC can only perform mesh reconstruction. We texture our parameterized meshes with a checkerboard pattern to evaluate the quality of mesh parameterization. Our results show that our method generates a well-parameterized mesh with no overlap.

\begin{table}[b]
    \begin{center}\fontsize{8}{4}\selectfont
    \caption{Evaluation for the reconstruction of shap   e Spot, Bunny,and Duck, where our method have a comparable reconstruction ability to Marching Cube}\label{CDplane}
    \begin{tabular}{ccccc}
    \toprule  
    Methods & \multicolumn{1}{c}\textbf{MC}   &  \multicolumn{3}{c}{\textbf{Ours}}\\
    \toprule  
    Metrics & CD$\downarrow$  & NC$\uparrow$ & CD$\downarrow$  & NC$\uparrow$\\
    \midrule  
    Spot & 1.14  & 0.46 & 1.13 & 0.54 \\
    Bunny& 1.49 & 0.49 & 1.47 & 0.50 \\
    Duck & 1.47  & 0.50 & 1.46 & 0.50 \\

    \bottomrule 
    \end{tabular}
    \end{center}
    \label{tablequan}
\end{table}

\subsection{Quantitative results}

We further validate the effectiveness of our method by experimenting on additional data from the ABC test set. We selected genus 0 shapes and successfully generated parameterized remeshing for all 51 shapes, achieving a 100\% success rate. The quantitative results are presented in Table~\ref{tablequan}, where CD (Chamfer Distance) measures the difference between the reconstructed and ground truth meshes, and NC (Normal Consistency) evaluates the surface normals. These values were computed for each shape and averaged, demonstrating the effectiveness and robustness of our method across multiple shapes.


\begin{table}[b]
    \begin{center}\fontsize{8}{5}\selectfont
    \caption{Quantative results for surface reconstruction from the ABC data set, where Chamfer Distance (CD) evaluates the overall quality of a reconstructed mesh, and Normal Consistency (NC) measures the accuracy of the surface normals.}\label{CDplane}
    \begin{tabular}{cccc}
    \toprule  
    Metrics & CD$\downarrow$  & NC$\uparrow$ \\
    \midrule  
    Marching Cube & 69.52  & 0.46 \\
    Our Method & 69.36 & 0.46 \\
    \bottomrule 
    \end{tabular}
    \end{center}
\end{table}
\subsection{Comparison with previous work} 
Since no prior work has reconstructed differentiable surfaces from implicit neural representations, direct comparisons with traditional mesh reconstruction or parameterization algorithms are not feasible. Comparing our method with state-of-the-art isosurface extraction techniques like FlexiCube~\cite{flexicube} and Arcs~\cite{24arcs} is also challenging due to fundamental differences in methodology and scope. Unlike these methods, which focus solely on mesh generation, our approach integrates both reconstruction and parameterization, complicating direct comparisons. Additionally, traditional parameterization techniques rely on explicit mesh inputs, whereas our method works with implicit representations, further complicating comparisons.

By integrating reconstruction and parameterization into a single step, we preserve the differentiability of the original shape and achieve a completely smooth parameterization. This seamless integration is not possible when reconstruction and parameterization are performed sequentially using discrete meshes.

Despite these methodological differences, our approach offers a theoretical advantage in speed. It efficiently scans the curved surface to precisely determine vertex positions on the target mesh, in stark contrast to cube methods like MC and Flexicube, which segment the space into cubes. This capability allows our method to process more rapidly and efficiently.

\section{Discussion and Conclusions}
We propose a novel shrinking method for reconstructing surfaces from deep signed distance fields. Compared to previous approaches, our method provides significant advantages by producing differentiable and parameterized surfaces, fully leveraging the benefits of continuous deep implicit representations. More importantly, this completes the final piece of the geometry generation process, ensuring that the entire pipeline remains differentiable when calculating the loss, rather than discretizing the network during the final geometric reconstruction. Additionally, our technique supports sphere parameterization. Looking ahead, we aim to extend our approach to more complex shapes with genus greater than zero, using implicit deformations between shapes.

Currently, our method lacks a mechanism to prevent parameterization distortion during surface reconstruction. To address this, we plan to incorporate a term based on the neural SDF's curvature and utilize the First and Second Fundamental Forms of the parameterization, minimizing distortion and improving the fidelity of the parameterization.



\bibliography{tracerefer}
\vspace{12pt}

\bibliographystyle{IEEEtran}

\end{document}